\documentstyle[aps,epsf,psfig]{revtex}
\begin{document}
\twocolumn[\hsize\textwidth\columnwidth\hsize\csname @twocolumnfalse\endcsname

\title{Theory of Excitons in Insulating Cu-Oxide Plane}
\author{F. C. Zhang, K. K. Ng}
\address{
Department of Physics, University of Cincinnati,
Cincinnati, Ohio 45221\\ 
}

\maketitle

\begin{abstract}
We use a local model to study the formation and the structure of the low 
energy charge transfer 
excitations in the insulating Cu-O$_2$ plane. The elementary excitation is a 
bound exciton of spin singlet, consisting of a Cu$^+$ and a neighboring spin 
singlet of Cu-O holes. The exciton can move through the lattice freely without
disturbing the antiferromagnetic spin background, in contrast to the single
hole motion. There are four eigen-modes of excitons with different symmetry. 
The $p$-wave-like exciton has a large dispersion width. The 
$s$-wave-like exciton mixes with p-state at finite momentum, and its 
dipole transition
intensity is strongly anisotropic. The model is in excellent agreement with
the electron energy loss spectra in the insulating Sr$_2$CuO$_2$Cl$_2$.

\noindent PACS numbers: 71.35.Cc, 74.72.Jt
\end{abstract}

\vskip2pc]
\newpage
\section{Introduction}
There have been extensive efforts in recent several years on the electronic
structure in the layered Cu-oxides, in the hope to reveal the mechanism for
the high $T_c$ superconductivity. It is now well established that the low 
energy physics of the undoped cuprates is well described by the spin-1/2
Heisenberg model in a square lattice, leading to an antiferromagnetic 
insulator. Two-dimensional $t-J$ model has been proposed and argued to describe 
the low energy physics for the cuprates \cite{Anderson,Zhang}, and it seems
the model explains many unusual properties of the high $T_c$ materials.

In this paper we present a theory for the
lower energy charge transfer excitations in the insulating Cu-oxide plane. Our
work was motivated by the angle-resolved electron energy loss spectra (EELS)
in the insulating Sr$_2$CuO$_2$Cl$_2$, where the optical allowed transition
shows a large energy dispersion width ~1.5 eV, and the optical forbidden 
transition was observed to be strongly anisotropic \cite{Wang}.

We use a local model to analyze the formation and the structure of the charge 
transfer excitation
between Cu-3$d_{x^2-y^2}$ and O-2$p\sigma$ in the Cu-oxide planes. The 
elementary charge transfer excitation is a bound exciton of spin singlet,
consisting of a Cu$^+$ (quasi-particle) and a neighboring spin singlet of 
Cu-O holes. The exciton can move through the lattice freely without disturbing
the antiferromagnetic (AF) spin background, in contrast to the single hole
motion. There are four exciton modes related to the four-fold rotation 
symmetry in the lattice, whose hopping matrices determine the dynamics and 
the symmetry of the charge transfer excitations. The $p$-wave like exciton has
a large energy dispersion width. The $s$-wave-like exciton mixes with the 
$p$-state
at finite {\bf k}, and its optical transition intensity is strongly 
anisotropic. The 
model is in excellent agreement with the recent electron energy loss spectra in
the insulating Sr$_2$CuO$_2$Cl$_2$. The theory also predicts a lower energy
optical forbidden mode, which should be observable in luminescence experiment
\cite{Wells}. Since the mass of the particle-hole pair is heavier than the
particle or the hole in a usual semiconductor or in other band-like insulator
materials \cite{Mattis}, the anomalous behavior of the dispersion widths 
between the single hole and the exciton in cuprates provides further evidence
for the strong correlation in these compounds. A part of the results in this
paper was briefly reported in a joined article with the EELS experiments 
\cite{Wang}. The paper is organized as follows. In section \ref{second},
 we examine the 
formation and the structure of small excitons in the insulating cuprates. The
dynamics of the excitons is derived in section \ref{third}. The exciton 
eigen-modes and
the relevance to the charge transfer gap and to the EELS are described in 
section 
\ref{forth}. Several remarks and discussions are made in section \ref{fifth}.

\section{Structure of small excitons}\label{second}
We consider a Cu-oxide plane, where the vacuum is the filled shell states of
Cu$^+$ and O$^{2-}$. In the undoped case, there is one hole in average in each
unit cell of CuO$_2$, and the hole primarily resides on the Cu site and is of 
$d_{x^2-y^2}$ symmetry. Because of the strong on-site Coulomb repulsion between
the Cu holes, the ground state is a spin-1/2 antiferromagnet of Cu$^{2+}$. To
study the low energy charge excitations of a particle-hole pair, we shall focus
on the transition from the Cu-3$d_{x^2-y^2}$ to O-2$p\sigma$ states. This
transition is of particular interest due to the relatively small atomic energy
 difference and due to 
the large orbital overlap between the two states hence the strong intensity in
experiments. The measured optical gap \cite{Uchida} is related to this 
transition.

Let us start with a local pair of particle and hole and examine the formation
and the structure of the exciton. As illustrated in Fig.~\ref{excitons}, the 
quasi-particle
is at a vacant Cu-site (Cu$^+$), and the quasi-hole is at the O-site. There are
two important physical effects for the pair. The first is the Coulomb 
attraction
between the particle and the hole, which should be considerably large in the
insulating compounds where the screening effect is weak. It is this attraction
to bind the pair to form an exciton. For simplicity we shall assume the 
attractive potential to be short-ranged, which is $-E_c<0$ when the particle 
and
hole are nearest neighbors (n.n.), and is zero otherwise. We expect that the
more realistic potentials will not change the basic physics as long as the
n.n. attraction dominates. The second is the strong Cu-O hybridization which
binds a hole on each square of O-atoms to the central Cu-hole to form a spin
singlet \cite{Zhang}. Let us denote this binding energy to be $E_s$. As a
compromise of the above two effects, we expect a spin singlet with more 
weight of the O-hole on the site near the Cu$^+$. We can make the above 
argument more quantitatively using variation approach. Let the O-hole state be

\begin{equation}
\Phi=\alpha P_{-\hat{x}/2}+\beta \left(-P_{\hat{x}/2}+P_{\hat{y}/2}-
P_{-\hat{y}/2} \right),
\end{equation}

\noindent where $P$ is the atomic O-hole orbital, whose sub-index denotes the
atomic position relative to their central Cu-hole (lattice constant=1), and 
P$_{-\hat{x}/2}$ is
the one close to the Cu$^+$. The total binding energy of the exciton, $E_b$,
takes maximum when $\alpha^2=(1+1/\sqrt{1+3c})/2$, with $c=E_s^2/(2E_c-E_s)^2$.
Typically, $\alpha^2=0.5$ or 0.75 for $E_c/E_s$=0.5 or 1. From the above 
analysis we see that the exciton in cuprates is a complex object, consisting
of a Cu$^+$ and a nearby Cu-O singlet. The spin triplet exciton energy has 
a binding energy of $E_c$, smaller than the binding energy $E_b$ of the
spin singlet exciton, and will not be considered in this paper. 

The parameters $E_c$ and $E_s$ in cuprates should be within certain
regions so that the system is stable against the excitonic excitations. Let 
$\epsilon_p$ be the atomic energy difference between the Cu- and O-hole states,
the stability condition in the local limit is $E_b < \epsilon_p$. If we further
consider an extra hole (primarily residing on O-site) in the half-filled 
insulating Cu-oxides, the model system should be stable against excitonic
excitation (one of the Cu-hole moves to the oxygen). This implies 
$2 E_b < \epsilon_p+E_s$. Our discussion in this paper will be limited to
the above physical parameter region.

There are four spin singlet exciton states for a fixed quasi-particle at 
{\bf R}, 
denoted by $\tau({\bf R})$, where $\tau$ is the coordinator of the Cu-site in 
the singlet relative to {\bf R}. $\tau=\hat{x},~-\hat{x},~\hat{y}$, and 
$-\hat{y}$. These will be briefly denoted by $x,~\bar{x},~y,$ and $\bar{y}$
respectively (see Fig.~\ref{excitons}).

\section{Dynamics of excitons}\label{third}
The exciton can move through the lattice due to the n.n. Cu-O hopping $t_{pd}$
and the O-O direct hopping $t_{pp}$. The most important feature of the exciton
dynamics is that the exciton motion does not disturb the AF spin background.
This is because each exciton involves two neighboring Cu-sites, both are
spinless. As far as spins are concerned, the spin singlet exciton is similar
to a pair of bound n.n. holes, whose motion does not disturb the AF spin
background. 
This property should be compared with the single hole dynamics in Cu-oxide, 
where the AF spin background is strongly disturbed, and the hole dispersion is
given by the Cu-Cu spin exchange interaction $J$ instead of the hole hopping
integral $t$, as shown theoretically for the $t-J$ model \cite{Liu} and 
observed in the recent angle-resolved photoemission experiments \cite{Wells}

We now study the exciton motion quantitatively. A general hopping process of
the exciton in the coordinator space may be represented by $\tau({\bf R})
\rightarrow \tau'({\bf R}')$ with the hopping integral $t_{\tau\tau'}({\bf R}
-{\bf R}')$. Because the AF spin background is unchanged \cite{hopping}, the 
exciton motion is equivalent to that for a free particle. Let $|GS\rangle$ be 
the AF ground state, and $\gamma^\dagger_{\tau}({\bf R})$ is an operator to 
create an exciton $\tau({\bf R})$ in the ground state, so that 
 $\gamma^\dagger_{\tau}({\bf R})|GS \rangle$ is  the state
of exciton $\tau({\bf R})$ in an otherwise AF spin background. The exciton
hopping is described by an effective Hamiltonian $H_{eff}$, which acts on the
Hilbert space of the single exciton states, 
\begin{equation}
H_{eff}\gamma^\dagger_{\tau}({\bf R})|GS\rangle=\sum_{{\bf R}', \tau'} 
t_{\tau\tau'}
({\bf R}-{\bf R}')\gamma^\dagger_{\tau'}({\bf R}')|GS\rangle. 
\end{equation}

\noindent It follows that

\begin{equation}
H_{eff}=\sum_{{\bf k}}\gamma^\dagger({\bf k})T({\bf k})\gamma({\bf k}),
\end{equation}
\noindent with $\gamma^\dagger({\bf k})=(\gamma^\dagger_x, 
\gamma^\dagger_{\bar{x}}, 
\gamma^\dagger_y, \gamma^\dagger_{\bar{y}})$. $T({\bf k})$ is a 4 by 4 matrix,
which determines the exciton dynamics, and $T_{\tau\tau'}({\bf k})=\sum_{\bf r}
t_{\tau\tau'}({\bf r})e^{i{\bf k\cdot r}}/N$, with $N$ the number of Cu-sites.
Note that {\bf k} is in the
whole first Brillouin zone although there are two sub-lattices in the ground 
state.
The matrix is hermitian, $T_{\tau\tau'}({\bf k})=T_{\tau'\tau}^*({\bf k})$.
Because $t_{\tau\tau'}({\bf r})$ is real, we have $T_{\tau\tau'}({\bf k})=
T_{\tau\tau'}^*(-{\bf k})$. The Cu-O plane has certain symmetries. Taking into
account of the Cu and O-hole orbital symmetries, we have 
$T_{\tau\tau'}({\bf k})=T_{\bar{\tau}\bar{\tau}'}(-{\bf k})$ from the parity
invariance, $T_{x\bar{y}}(k_x,k_y)=-T_{xy}(k_x,-k_y)$ from the inversion 
symmetry with respect to the x-axis, and 
$T_{x\bar{x}}(k_x,k_y)=T_{y\bar{y}}(-k_y,
k_x)$ due to the four-fold rotational invariance.

To pursue the theory further, we estimate the hopping integral $t_{\tau\tau'}
({\bf r})$ within the atomic limit and treat $t_{pd}$ and $t_{pp}$ 
perturbatively. We shall consider the limit $\alpha=1$ in Eq.(1) for simplicity
and for the purpose of illustration. Up to the third order in perturbation in
$t_{pd}$ or $t_{pp}$, there are only four non-zero independent integrals:

\begin{eqnarray}
t_1&\equiv& t_{xy}(0)=\frac{1}{2}(t_{pp}-t^2_{pd}/\epsilon_p),\nonumber \\ 
t_2&\equiv& t_{xy}(\hat{x})=t_{pp}t^2_{pd}/(2\epsilon_p E_c),\nonumber \\
t_3&\equiv& -t_{x\bar{x}}(\hat{x})=t^2_{pd}[\epsilon_p^{-1}-(\epsilon_p+U_{pp}
)^{-1}],\nonumber \\
t_4&\equiv& t_{x\bar{x}}(0)=t^2_{pd}/(2\epsilon_p).
\label{integrals}
\end{eqnarray}

\noindent A diagrammatic illustration for these hopping processes are shown in
Fig.~\ref{hoppings}. In Eq.~(\ref{integrals}), U$_{pp}$ is the on-site O-hole 
Coulomb
repulsion. We have kept the third order term, $t_2$, because $E_c$ is small
compared to $\epsilon_p$. From Eq.~(\ref{integrals}), $t_3>t_4$. Since
$t_{pp}>0$ for the O-oxygen hole hopping ,
$t_2$, $t_3$, and $t_4$ are all positive. Using these $t's$ and the symmetries
discussed above, we obtain 

\begin{equation}
T({\bf k})=\left( \begin{array}{cccc}
0 & a(k_x) & b({\bf k}) & \bar{b}({\bf k}) \\
a^*(k_x) & 0 & \bar{b}^*({\bf k}) & b^*({\bf k}) \\
b^*({\bf k}) & \bar{b}({\bf k}) & 0 & a(k_y) \\
\bar{b}^*({\bf k}) & b({\bf k}) & a^*(k_y) & 0 
\end{array} \right),
\label{matrix}
\end{equation}

\noindent where $\bar{b}(k_x,k_y)=-b(k_x,-k_y)$, and 

\begin{eqnarray}
a(k_x)&=&t_4-t_3 e^{ik_x},\nonumber \\
b({\bf k})&=&t_1+t_2(e^{ik_x}+e^{-ik_y}).
\end{eqnarray}

\section{Exciton eigen-modes}\label{forth}
$H_{eff}$ can be diagonalized to obtain the exciton solutions. For {\bf k} 
along [11] direction, the analytical solutions are particularly simple and are
listed in Table I. At {\bf k}=0, the four exciton eigenstates have well 
defined local symmetry \cite{Rice}, describing the uniformly moving molecular
states. They are illustrated in Fig.~\ref{symmetry}. The $s$-wave and
$d$-wave states are given by 

\begin{eqnarray}
|s\rangle &=&(-\frac{1}{2},\frac{1}{2},-\frac{1}{2},\frac{1}{2}),\nonumber \\
|d\rangle &=&(-\frac{1}{2},\frac{1}{2},\frac{1}{2},-\frac{1}{2}), 
\label{sd-wave}
\end{eqnarray}

\noindent where the four components represent
the amplitudes of the excitons $\gamma_x, \gamma_{\bar{x}}, \gamma_y, 
\gamma_{\bar{y}}$ respectively. The energies for $s$- and $d$-states are 
$2b_0-a_0$, and 
$-2b_0-a_0$ respectively, where $a_0$ and $b_0$ are their values at 
${\bf k}=0$, and $a_0=t_4-t_3>0$, $b_0=t_1+t_2$.  The two $p$-wave states are 
degenerate and the energy is $-a_0$. In
the parameter region $-b_0<a_0<b_0$ and $b_0>0$ (suitable for Cu-oxide, see 
Fig.~\ref{spects} caption), the
$s$-wave state has the highest energy \cite{molecular}. The optical 
experiment measures the dipole active mode at $k$=0. Because of the
$d$-symmetry of the Cu-hole, both the $s$- and $d$-wave modes are optically
forbidden. The $p$-wave modes are optically active, and the charge transfer gap
is given by 

\begin{equation}
\Delta=\epsilon_p-E_b-a_0.
\end{equation}

For finite $k$ along an arbitrary direction $\hat{k}$, the eigenstate of
the exciton is a linear combination of the molecular states. We shall denote
the four exciton modes by $P_1$, $P_2$, $S$, and $D$, according to their 
symmetries $p_1$, $p_2$, $s$, $d_{x^2-y^2}$ in the limit $k\rightarrow 0$. For
the two $p$-states, we define a dipole active mode $|p_1\rangle$ and a dipole
inactive mode $|p_2\rangle$ as

\begin{eqnarray}
|p_1\rangle &=&(-k_x,-k_x,k_y,k_y)/(\sqrt{2}k), \nonumber \\
|p_2\rangle &=&(k_y,k_y,k_x,k_x)/(\sqrt{2}k).
\label{p-wave}
\end{eqnarray}

\noindent The exciton dispersions are plotted 
in Fig.~\ref{spects}.
The parameters are chosen \cite{transition} to fit the measured EELS
\cite{Wang}. The dispersions are anisotropic. The $E-{\bf k}$ relation of the
$P_1$ mode is monotonic, and the energy is peaked at ${\bf k}=(\pi,\pi)$. The 
dispersion width is $2t_3\sim 2t^2_{pd}/\epsilon_p$, which is a large energy
scale, order of the hole hopping
integral in cuprates, but much larger than the dispersion width of the single
hole motion in the AF background which is $\sim J$ \cite{Liu}. This result is 
consistent with the EELS \cite{Wang}. The dispersion along [10] direction is
much flatter, also consistent with the EELS. As we can see from 
Fig.~\ref{spects}, the lowest energy mode is the $D$-mode. This is in agreement
with the week coupling method of Littlewood, Varma, and Schmitt-Rink 
\cite{Littlewood}.

In the EELS, an incident electron is inelastically scattered to create a pair
of electron-hole with energy and crystal momentum ($\omega$, {\bf k}). The EELS
directly measures Im[$-1/\epsilon({\bf k}$, $\omega$)], with 
$\epsilon$({\bf k}, $\omega$) the dielectric function, hence probes the 
dispersion and the 
symmetry of the charge transfer excitation \cite{Schnatterly}. For an 
exciton, the intensity in the imaginary part of $\epsilon$({\bf k}, $\omega$)
around a pole is,

\begin{equation}
I=const*k^{-2}\left|\langle \Psi_{ex}|e^{i{\bf k}\cdot {\bf r}}|GS\rangle
\right|^2,
\end{equation}

\noindent where $\Psi_{ex}$ is the exciton state. For small {\bf k}, the most
dipole active excitations in the EELS can be identified to be the $P_1$ mode.
The dispersion obtained from the EELS \cite{Wang}
compares well with the theory. As $k$ increases, the $S$- and $D$-mode 
excitons gradually become EELS active because of the mixing of the 
$|p_1\rangle$ molecular state and because of the quadrapole contribution. It
is difficult to make a quantitative comparison between the dipole and 
quadrupole contributions in the $S$- or $D$-mode. However, if the radius of
the Cu-3$d$ state is much smaller than the lattice constant, the dipole
contribution is more important. We expect this to be the case in cuprates. 
Within the dipole approximation, we have

\begin{equation}
I=I_0\left|\langle\Psi_{ex}|p_1\rangle\right|^2.
\label{inten}
\end{equation}

\noindent where $I_0$ is the dipole active intensity of $P_1$-mode at 
{\bf k}=0. Note the total dipole active intensity of the four modes is 
conserved to be $I_0$. In Fig.~\ref{intys}, we show the calculated 
intensity $I$ of the
excitons in Eq.~(\ref{inten}) as functions of {\bf k}. Along [10] direction,
as $k$ increases, the intensity of the $S$-mode increases rapidly, while the 
intensity of the $P_1$-mode decreases. The intensity along [11] is quite flat
for the $P_1$-mode and remains to be zero for the $S$-mode. This anisotropy is
 due to the symmetry of the exciton states. The $S$-mode eigenstate along [11]
contains only $s$ and $p_2$ states, both are dipole inactive. Furthermore, the
transition matrix from the quadrupole contribution is proportional to 
$\langle \Psi_{ex}|e^{i{\bf k}\cdot {\bf r}}|GS\rangle$, which is ~
$(k^2_x-k^2_y)$ for the $s$-state \cite{similar}. Therefore, the intensity for
the $S$-mode along [11] vanishes up to the quadrupole order. These features
are in excellent agreement with the EELS experiment \cite{Wang}, where an 
optical 
forbidden transition was observed along [10] but not along [11] direction, and
the intensity of the optical active transition along [10] direction is observed
in the experiment to decrease as {\bf k} increase. 

The $D$-mode exciton, expected from the theory, is another optical forbidden
state. The current EELS has not revealed this mode, probably due to the limit
of the experiment resolution. Further spectroscopy measurements are needed to 
verify
this $d$-wave-like state. This mode can be active in the phonon-assisted 
optical process, and should be observable in the luminescence experiment. A
recent optical absorption measurement indicates a very weak absorption state
at about 0.5 eV in the undoped cuprates \cite{similar}. The $d$-symmetry 
state is an alternative to the magnon state or crystal field exciton proposed
earlier for this weak absorption where the phonons could be involved.

\section{Discussions}\label{fifth}
In this paper we discussed small excitons in the insulating Cu-oxide planes. 
These are the charge transfer excitations, which are separated from the lower
energy excitations by an energy order of $\epsilon_p$. The lower energy 
excitations are the magnons described by spin -1/2 Heisenberg model at 
half-filled. In the doped case, the lower energy charge excitations are given
by the hole motion in the AF spin background described by the 2-dimensional
 $t-J$ model.

We have used a local model to describe the excitons in Cu-oxide planes. If we 
include the individual kinetic energies of the quasi-particle and the 
quasi-hole, the excitons should have spatial extensions. The size of the 
exciton is determined by the balance of the kinetic energies and the Coulomb
attraction $E_c$. This spatial extension will lower the exciton energy. The
exciton motion will be more complicated, and will contain an incoherent part
disturbing the AF background. The large dispersion width observed in the EELS
may be viewed as an experimental indication that the spatial extension of the
exciton does not play an important role and the small exciton proposed here
may be a good approximation.

We have so far not considered the electron-hole (e-h) continuum, which starts
at energy $\Delta_{e-h}=\epsilon_p-E_s+E_{Kin}$, where $E_{Kin}<0$ is the 
lowest kinetic energy of the independent Cu$^+$ and the formal Cu$^{3+}$. For 
the cuprates, $|E_{Kin}|\sim 1-2$~eV. We expect the e-h continuum to start 
above the {\bf k}=0 $p_1$ mode, and the exciton spectra extend into the e-h 
continuum due to the large spectral dispersion. The states within the 
continuum region will then be damped, but can exist as resonant states 
contributing to the EELS. Since the contributions to the EELS from the e-h
continuum have much less {\bf k}-dependence, the excitons become the dominant
source of the {\bf k}-dependent spectra in the EELS.

Our model applies to the insulating
phase. In the metallic phase the Coulomb attraction between the quasi-hole
and quasi-particle is substantially reduced due to the metallic screening, and
is too weak to bind a pair.

In conclusion, we have studied a local exciton model for the insulating 
Cu-oxides. The exciton moves through the lattice almost freely in the AF
spin background. The model is in excellent agreement with the recent EELS in 
the insulating Sr$_2$CuO$_2$Cl$_2$.

\section{Acknowledgment}
We wish to thank Y.Y. Wang and M.V. Klein for numerous discussions on both
experimental and theoretical aspects related to the present work. We would
also like to thank T.M. Rice and C.M. Varma useful discussions.


\newpage
\begin{figure}[htbp]
\caption{Structure of local excitons in insulating Cu-O planes. The open 
circles represent O-atoms, the solid circles represent Cu-atoms. (a) exciton
$x({\bf R})$: the quasi-particle is at the vacant Cu-site {\bf R}, and the
quasi-hole is on the square of O-atoms with more weight at the left atom and
forms a spin singlet with the central Cu-hole at position ${\bf R}+\hat{x}$.
Not shown are a single hole on all other Cu-sites. (b) a simpler 
representation of (a). (c), excitons $\bar{x}({\bf R})$, $y({\bf R})$ and 
$\bar{y}({\bf R})$, with the spin singlet at different positions relative to
{\bf R}.} 
\label{excitons} 
\end{figure} 

\begin{figure}[htbp]
\caption{A list of five most important hopping processes from $x({\bf R})$ to
$\bar{x}({\bf R}')$ and $y({\bf R}')$. The other hoppings can be obtained by
symmetry. The corresponding effective hopping integrals are given by the 
Eq.~(\protect\ref{integrals}).}
\label{hoppings} 
\end{figure} 

\begin{figure}[htbp]
\caption{Symmetries of the four molecular exciton wavefunctions, $s$- and 
$d$-waves in Eq.~(\protect\ref{sd-wave}), and the two $p$-waves in 
Eq.~(\protect\ref{p-wave})
with $k_x=k_y$. The center $d$-wave represents Cu$^+$, a vacancy at Cu-site. 
The 
oxygen $2p_x(2p_y)$ hole wavefunctions are represented by their orbitals. Only
charge degrees of freedom are shown. The implicit spins are the same as in
Fig.~\protect\ref{excitons}.}
\label{symmetry} 
\end{figure}

\begin{figure}[htbp]
\caption{Calculated exciton energies E (eV) as a function of 
{\bf k}$=(k_x, k_y)$, for the parameters 
$t_1=0.4$, $t_2=0.126$, $t_3=0.85$, $t_4=0.65$~eV. The energy is measured
relative to the $p_1$-mode at {\bf k}=0. The thick solid lines show the
dispersions along [10] direction. The dispersions of the $P$-wave mode
are compared very well with the EELS along [10] and [11] dispersions. See
Ref. 3 for the detailed comparison.}
\label{spects} 
\end{figure}

\begin{figure}[htbp]
\caption{Calculated intensities I (arb. unit) of the exciton transitions as 
function of 
{\bf k}, normalized to $I_0$. The parameters are as same as those in Fig. 4.
The thick solid lines show the intensities along [10] direction. The  
$P_2$-mode is not shown because it is dipole inactive and has no
contribution to the intenisty for all {\bf k}. 
The theory is compared well with the EELS along [10] and [11] directions. The
intensities for the $S$-mode at finite {\bf k} along [10] were observed in
the EELS, see Ref. 3.}
\label{intys} 
\end{figure}

\newpage
\vbox{
\begin{table}[b]
\begin{tabular}{|crcc|}
Mode  & energy  & eigenstate & symmetry at ${\bf k}\rightarrow 0$  \\ \hline
$P_1$ & $-b+|a-\bar{b}|$ & $(-\eta_{-},-1/2,\eta_{-},1/2)$ & $p_1$ \\ \hline
$P_2$ & $b-|a+\bar{b}|$  & $(-\eta_{+},1/2,-\eta_{+},1/2)$ & $p_2$ \\ \hline
$S$   & $b+|a+\bar{b}|$  & $(\eta_{+},1/2,\eta_{+},1/2)$   & $s$   \\ \hline
$D$   & $-b-|a-\bar{b}|$ & $(\eta_{-},-1/2,-\eta_{-},1/2)$ & $d_{x^2-y^2}$ \\
\end{tabular}
\vspace{0.5cm}
\caption{Solutions of the four exciton modes for {\bf k} along [11] direction.
The {\bf k}-dependence in $a$, $b$ and $\bar{b}$ is implied, see 
Eq.(\protect\ref{matrix}). $\eta_{\pm}=\frac{1}{2}(a\pm \bar{b})/|a\pm 
\bar{b}|$. The last column applies to the region $(-b_0<a_0<b_0)$ suitable to 
cuprates, and $a_0$, $b_0$ are their values at $k=0$.} 
\label{solutions}
\end{table}
}


\begin{references}
\bibitem{molecular}This is consistent with the local molecular picture. The
intra-molecular energies of the exciton states are 2$t_1$ for the $s$-wave,
$-2t_1$ for the $d$-wave, and zero for the two $p$-wave states. This leads to
the same conclusion that the S-mode has the highest energy provided $t_1>0$.
\bibitem{transition}There are two nearby peaks in the optical forbidden 
transitions observed in the EELS in Ref. 3. Here we assume 
the lower energy 
state to be the S-mode exciton discussed in the text. A smaller $t_1$ gives a 
sharper drop in the intensity of the $P_1$ mode along [10] direction.
\end{references}
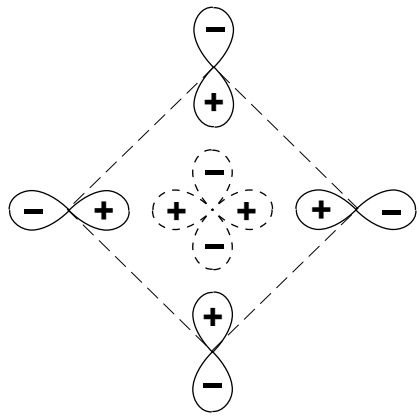
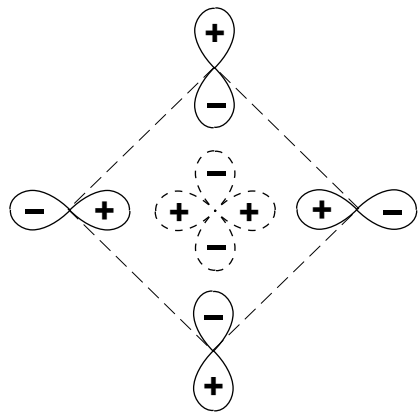
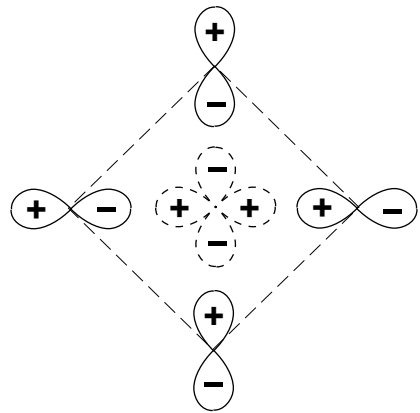
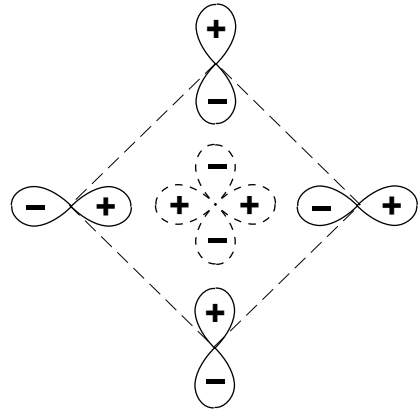
\end{document}